\documentclass[epj]{svjour}
\usepackage{graphics}

\begin{document}

\title{Coherent phase control in ionization of Magnesium by a bichromatic
laser field of frequencies \( \omega  \) and 2\( \omega  \) }

\author{Gabriela Buic\u{a}-Zloh\inst{1,}\thanks
{permanent address: Institute for Space Sciences, P.O. Box MG-23, Ro 77125,
Bucharest-M\u{a}gurele, Romania} and
 L. A. A. Nikolopoulos\inst{1,}\inst{2}
 }

\institute{Institute of Electronic Structure and Laser, FORTH P.O. Box 1527, 
 Heraklion-Crete 71110, Greece \and
Dept. of Telecommunications Science, Univ. of Peloponnese, GR-221 00, Tripolis, Greece}

\abstract{
We have investigated the coherent phase control
on the \( 3p^{2} \) autoionizing state (AIS)
resonantly coupled with the ground state for Mg through a two- and
a four-photon transition simultaneously, using a bichromatic linearly
polarized laser field. The frequency is chosen such that the lasers
are tunable around resonance with the transition
 \( 3s^{2}(^{1}S^e)\rightarrow 3p^{2}(^{1}S^e) \),
which implies \( \omega _{f}=2.11 \) eV and \( \omega _{h}=4.22 \)
eV. We are interested in the modification of autoionizing (AI) line
shape through the relative phase and laser intensities. A strong phase
dependence on the total ionization yield and ionization rate is found.
We also performed a time-dependent calculation which takes into consideration
all the resonant states of the process. 
\PACS{32.80.Qk Coherent control of atomic interactions with photons -
 32.80.Rm Multiphoton ionization and excitation to highly excited states (e.g., Rydberg states) -
32.80.Dz Autoionization }
}
\authorrunning
\titlerunning
\maketitle
\date
\section{Introduction}

This paper treats a particular aspect of the control of photoabsorption,
 and in particular,  ionization through the relative phase of two electromagnetic fields 
acting simultaneously on an atomic system.  The aspect in question is the possibility of
 altering the line shape of an autoionizing resonance through such phase control.  
The general features of phase control in photoabsorption have been discussed in the
literature extensively \cite{shap,chen1,chen2,cara,cav,taka2,milo}. 
 The possibility of altering an autoionizing
 line shape has also been discussed  \cite{lyr} to some extent in theoretical work where
 it has been
 shown that one particularly intriguing feature, depending on parameters,
is the possibility of turning off the transition to the "discrete"  or the continuum
 part of the resonance \cite{taka1,taka3}.  To the best of our knowledge, experimental observation of 
such an effect has not been recorded, one of the reasons perhaps being the unavailability
 of radiation sources of intensity and frequency suitable for convenient atomic species and resonances.
 
The case studied in this paper aims at presenting a quantitative analysis of a situation 
 possibly convenient for existing laser sources whose potential in phase control has been tested 
experimentally. We have thus undertaken the study of the line shape of the
$3p^2$ ($^1S^e$) autoionizing resonance of atomic Magnesium under the simultaneous excitation by 
four- and two-photon  transitions whose relative phase is varied.  This case  \cite{pap} is somewhat
different from, and theoretically a bit more demanding, than the more standard scheme of one- and
three-photon transitions, which in any case would not be applicable here owing to the
 parity of the resonance which is the same as that of the ground state.  
In addition, we have found that bound states in near resonance with
 one- or  three-photon transitions introduce distortions in the wings
 of the  autoionizing resonance;
 an effect not a prior obvious but which has to be taken into account for a realistic
 assessment of the observability of the desired feature.  Needles to say that a realistic atomic
structure and transitions calculation has been necessary, whose details are discussed
 in the sections that follow.
\\ 
Lyras and Bachau  \cite{lyr}  studied the phase control in
two- and four-photon ionization of the Magnesium atom in a bichromatic field of 
frequencies $\omega$ and $3\omega$, in the vicinity of an autoionizing
resonance $ ^{1}D^e $
lying above the first two ionization thresholds. Moreover, they accomplished
a full perturbative, time-independent phase control calculation in Mg by
interfering three- and one-photon transitions to a single continuum
channel $ ^{1}P^o $. 
They studied  the dependence of the ionization rate as a function
of the relative phase for different laser intensities, and  noticed
that  the presence of an intermediate resonance may enhance the
overall ionization signal.
\\
 Kylstra {\it et al.} \cite{kylstra:1998}
have performed a non-perturbative ab-initio one- and two-color  calculation of
the multiphoton ionization of Magnesium, where the laser frequencies are chosen such
that the initial state of the atom is resonantly coupled to the autoionizing
$3p^2$, $3s3d$, and $4s5s$ resonances of the atom. Using  R-matrix Floquet 
theory, they studied single photon ionization from the ground state $3s^2$ and the $3s3p$ 
excited state of Magnesium in the vicinity of the $3p^2$  AI
resonance at  non-perturbative laser intensities.
A second low-frequency laser couples the $3p^2$ and $3s3d$ AI
states. This resonant system was studied experimentally and
theoretically in the perturbative domain.  Kylstra {\it et al.} also  investigated
 the harmonic intensity regime where the process  is no longer perturbative.

In section 2 we present the basic framework related to the
 study of the phase control in the vicinity of an AI state.
In section 3 we briefly give the ab-initio theoretical approach
 for the description of the Magnesium atomic structure. Further
 details can be found in \cite{chang:1993a,nikolopoulos:2003a,lambropoulos:1998}.
 Finally, in section 4 we present our results.

\section{Basic Equations}

We study the multiphoton ionization of Mg using a realistic atomic
model which describes the time-evolution of the system, investigating
the phase coherent effect on the \( 3p^{2} \) (\( ^{1}S^e \)) autoionizing state
resonantly coupled to the ground state of Mg, simultaneously through
two- and four-photon transitions, in the weak field regime. 
The laser frequency is such that the ground state of the atom is near
resonance with the \( 3s3p\) \( (^{1}P^{o } )\) ($E_{3s3p} \approx$ 4.34 eV)
 state of the Mg atom due to its second harmonic, and with 
\( 3s4p\) \( (^{1}P^{o } )\) ($E_{3s4p} \approx$ 6.11 eV) 
due to three-photon absorption from the fundamental.  

We consider the Magnesium atom consisting of a ground state
\( |g>\equiv |3s^{2} \) \( ^{1}S^e> \), an autoionizing state 
\( |a>\equiv |3p^{2} \) \( ^{1}S^e> \), and one  continuum corresponding to 
three different values of angular momentum
\( |c_{1}>\equiv |3s\epsilon s\, \, ^{1}S^e> \), 
\( |c_{2}>=|3s\epsilon d\, \, ^{1}D^e> \),
and \( |c_{3}>=|3s\epsilon g\, \, ^{1}G^e> \). The AIS is modeled as
a discrete state embedded into the continuum and coupled to the \( ^{1}S^e \)
continuum through the configuration interaction.

\begin{figure}
\centerline{\resizebox*{8cm}{!}{\includegraphics{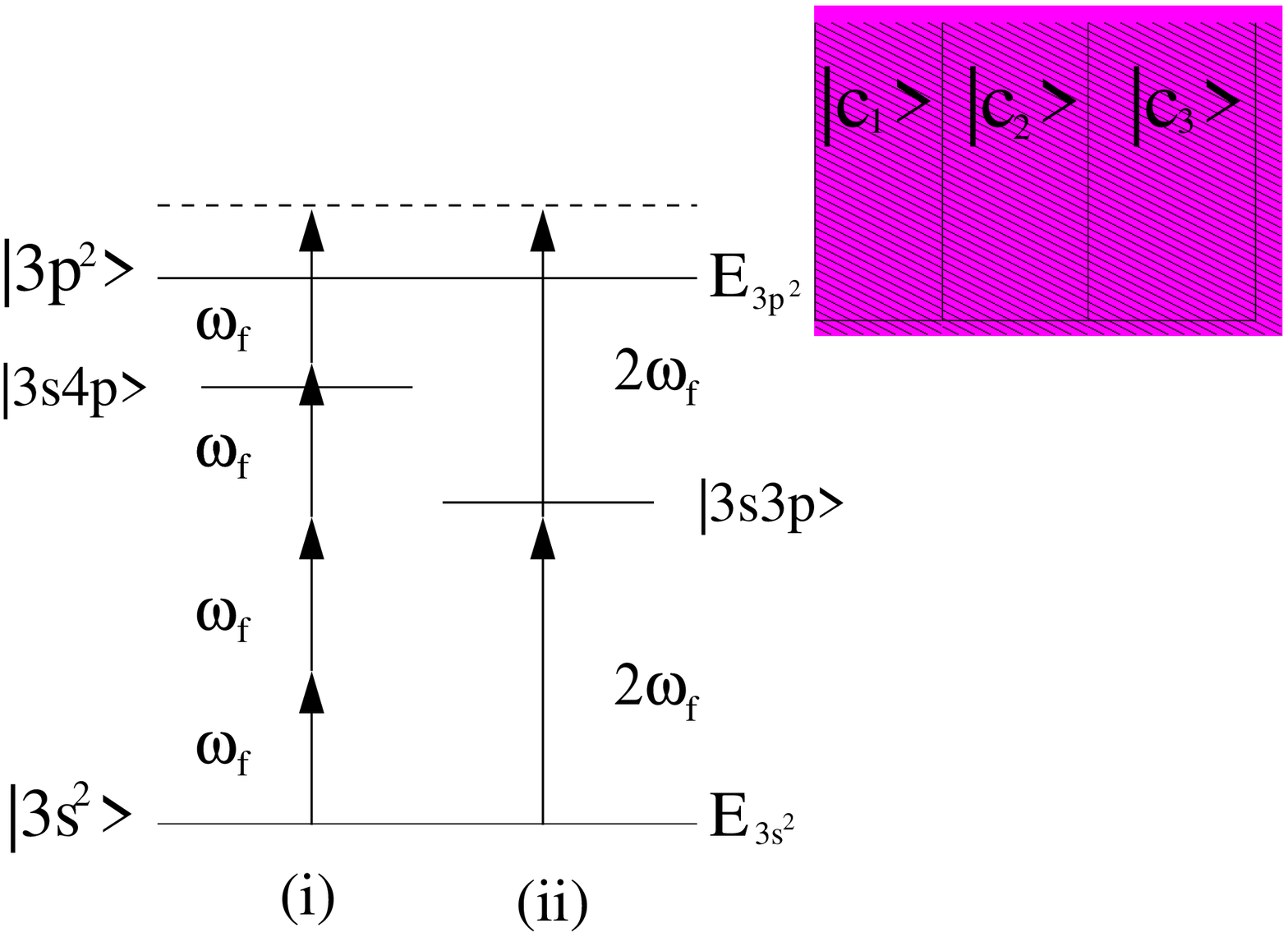}}} 
\caption{\label{fig:fig1}}\vspace*{1cm}
\end{figure}

In Figure \ref{fig:fig1} we show the energy diagram of the atomic system
interacting with the linearly polarized bichromatic field: 
\begin{equation}
\label{field}
{\bf{ E}}(t)={\bf E}_{f}(t)\exp {(i\omega _{f}t)}
+{\bf E}_{h}(t)\exp {(i\omega _{h}t+i\varphi )}+cc.,
\end{equation}
 consisting of a superposition between the fundamental with frequency
\( \omega _{f} \) and its second harmonic with frequency \( \omega _{h}=2\omega _{f} \).
The amplitudes of the two components of the electric field are \( {\cal E}_{f}(t) \),
\( {\cal E}_{h}(t) \), and \( \varphi  \) is the relative phase between
them. The continuum state of energy \( E_{g}+4\omega _{f} \) can
be reached by the two interfering paths shown in Figure \ref{fig:fig1},
namely, the four photon absorption from the fundamental electromagnetic
field (Figure \ref{fig:fig1}, path (i)), and the two-photon absorption
of its second harmonic (Figure \ref{fig:fig1}, path (ii)). 

There are a few other processes which  might affect the interference process, but for 
the intensities we consider in this paper they are not significant, therefore
we do not need to include them in the calculation. For instance, a three-photon transition
with two photons of the fundamental and one of the harmonic could lead
to the same final continuum energy. These processes, however, could only influence 
the background signal, and not the interference process, since the final continuum
state  belongs either to a $^{1}P^o$ or to a $^{1}F^o$ state.
We have estimated the transition amplitudes for these three-photon transitions and for values of the laser intensities  involved in our calculation, and we found out that the contribution to the ionization signal is about two orders of magnitude smaller than the contribution due to the processes presented in Figure \ref{fig:fig1}. One can therefore conclude that these three-photon processes do not affect the process studied here and  we  neglect them.
The dominant processes are decided by the value of the frequency and intensities
involved in the calculation.
For the same reason, Raman-type processes  are ignored due  to the fact that they
represent higher-order  processes with respect to the electric field.
 We have  used  atomic units throughout this work. 
The atomic unit used for the intensity of laser field is $I_0$=$14.037 
\times 10^{16} $ W/cm$^2$.

A four photon transition from the \( |3s^{2}\, \, ^{1}S^e> \) state
into the continuum leads to a final state containing \( |3s\epsilon s\, \, ^{1}S^e> \),
\( |3s\epsilon d\, \, ^{1}D^e> \), and \( |3s\epsilon g\, \, ^{1}G^e> \)
partial waves, while a two-photon transition into the continuum leads only to
\( |3s\epsilon s\, \, ^{1}S^e> \)
and \( |3s\epsilon d\, \, ^{1}D^e> \) partial waves. Thus, only the
transition amplitude to  the \( ^{1}S^e \) and \( ^{1}D^e \) continua is
modulated through the interference with the two-photon amplitude.
The transition amplitude to the \( ^{1}G^e \) continuum remains unaffected
by the quantum interference, and  therefore only contributes to the background
of the total ionization signal.

We consider the Schr\"{o}dinger equation: \begin{equation}
\label{Sch}
i\frac{\partial \Psi (t)}{\partial t}=\left[ H_{a}+D(t)\right] \Psi (t),
\end{equation}
 where \( H_{a} \) represents the atomic Hamiltonian.
The atomic system is treated as a two {\it active} electron system: the atomic
core (the nucleus and the \(10\) inner-shell electrons), and  two valence electrons.
More details about the atomic structure calculation are given in the next section.
 Within the semiclassical formalism, the interaction between the atom and the
bichromatic field is described in the length gauge, and in the dipole
approximation by \( D(t)=-({\bf r}_{1}+{\bf r}_{2})\cdot {\bf E}(t) \), where
 \({\bf r}_{1}\) and \({\bf r}_{2}\) represent the position vectors
of the two valence electrons.
In order to describe the dynamics of the system, the time dependent
wavefunction is expanded in terms of the complete set of states, and
then substituted into the Schr\"{o}dinger equation following the standard procedure \cite{lamb}. At \( t=0 \)
the system is assumed to be in the ground state \( |g> \), and for
time \( t>0 \) its wave function can be written as:
\begin{equation}
\label{wav}
|\psi (t)>=\tilde c_{g}(t)|g>+\sum ^{3}_{j=1}\int dE _{c_{j}}\, \,
 \tilde c_{j}(t)|c_{j}>,
\end{equation}
 where \(\tilde c_{g} \), and \( \tilde c_{j} \) are the probability amplitudes
of states \( |g> \) and \( |c_{j}> \), assumed to be eigenstates
of the atomic Hamiltonian \( H_{a} \) with the eigenenergies
\( E_{g} \) and \( E_{c_{j}} \), respectively. The substitution of the wave function
Eq. (\ref{wav}) into the Schr\"{o}dinger Eq. (\ref{Sch}) leads
to a set of differential equations for the amplitude coefficients.
We introduce the slowly varying coefficients \(c_g=\tilde c_g e^{i(E_g)t}\)
and \(c_j =\tilde c_j  e^{i(E_g+4\omega_f)t}\), and,
after the adiabatic elimination of the coefficients for the continuum
\( |c_{j}> \) in the differential equation for the coefficient of
the ground state, using the rotating wave approximation, we are left with one
independent equation (for the ground state), and three coupled differential
equations governing the time evolution of the probability amplitudes as such:
\begin{eqnarray}
&&i\frac{\partial c_{g}(t)}{\partial t}=\left[ S_{g}-\frac{i}{2}\gamma _{g}(\varphi ,t)\right] c_{g}(t), \label{sys1} \\
&&i\frac{\partial c_{j}(t,E_{c_{j}})}{\partial t}=
c_{j}(t,E_{c_{j}})\epsilon_{c_j}\label{sys2}\\ &&
+
\left[ D_{c_{j}g}^{(4)}(E_{c_{j}},t)
+e^{-2i\varphi }D_{c_{j}g}^{(2)}(E_{c_{j}},t)\right] c_{g}(t), \; j=1,\, 2, \nonumber \\
&&i\frac{\partial c_{3}(t,E_{c_{3}})}{\partial t}=
c_{3}(t,E_{c_{3}})\epsilon_{c_3}
+D_{c_{3}g}^{(4)}(E_{c_{3}},t)c_{g}(t), \label{sys3}
\end{eqnarray}
where \( \epsilon_{c_j}=E_{c_{j}}- E_g -4\omega_f \),  \(\; j=1,\, 2,\,3\).
\\
\noindent We used the identity:
\begin{equation}
\lim_{\eta\to 0^+}  \int\frac{f(x)}{x-x_0+i\eta}dx={\cal P} \int\frac{f(x)}{x-x_0}dx -i\pi f(x_0),
 \end{equation}
where ${\cal P}$ is the Cauchy principal value integral.
\noindent
$S_g $ represents the ground state AC-Stark shift
 induced by the laser field. 
Since we are working in the weak laser field regime, $S_g$
will be neglected.

 The quantities \( D_{c_{j}g}^{(2)} \) and \( D_{c_{j}g}^{(4)} \)
are the effective two- and four-photon transition amplitudes from 
 the ground state \( |g> \) to continuum \( |c_{j}> \):
\begin{eqnarray}
&&D_{c_{j}g}^{(2)}(E_{c_{j}},t) =  \sum _{n}\frac{M_{c_{j}n}^{(h)}\,\, M_{ng}^{(h)}}{\omega _{ng}-\omega _{h}}, \quad j=1,2 
\label{eq:cs2ph} \\
&&D_{c_{j}g}^{(4)}(E_{c_{j}},t)  =  \sum _{n,l,m}\frac{M_{c_{j}n}^{(f)}\, \, M_{nl}^{(f)}\, \, M_{lm}^{(f)}\, \, M_{mg}^{(f)}}{(\omega _{ng}-3\omega _{f})(\omega _{lg}-2\omega _{f})(\omega _{mg}-\omega _{f})},\nonumber 
\\&& j=1,2,3,
\label{eq:cs4ph} 
\end{eqnarray}
 where the one-photon transition amplitude between the
state \( |A> \) and \( |B> \) are written in the length gauge as 
\( M_{AB}^{(h)}(t)=-<A|({\bf r}_{1}+{\bf r}_{2})\cdot {\bf E}_{h}(t)|B> \),
and
 \( M_{AB}^{(f)}(t)=-<A|({\bf r}_{1}+{\bf r}_{2})\cdot {\bf E}_{f}(t)|B> \),
respectively, and \( \omega _{AB}=E_{A}-E_{B} \). 

\noindent The quantity \( \gamma _{g} \) represents the effective
total ionization width directly into the continuum, from the ground
state: 
\begin{equation}
\gamma _{g}(\varphi ,\, t)=2\pi \sum ^{2}_{j=1}\left| \bar{D}_{gc_{j}}^{(4)}(t)+\bar{D}_{gc_{j}}^{(2)}(t)e^{2i\varphi }\right| ^{2}+
2\pi \left|\bar{D}_{gc_{3}}^{(4)}(t)\right|^{2},
\end{equation}
 where the bar over the effective transition amplitudes
\( \bar{D}_{gc_{j}}^{(2)}\), and \( \bar{D}_{gc_{j}}^{(4)} \)
means that they have been calculated at \( E_{c_{j}}=E_{g}+4\omega _{f} \),
\( j=1,2,3 \). 

\noindent Firstly, we consider the ionization probability per
unit time \( P \), which is valid for an almost square pulse 
${\cal E}_f \equiv {\cal E}_f(t)$, and ${\cal  E}_h \equiv {\cal E}_h(t)$.
In this case, the photoionization line shape is simply obtained in
terms of the transition rate without any time-dependent calculations
by assuming \( \partial _{t}c_{j}(t)=0 \). The ionization rate is
given as the loss of population from the ground state: 
\begin{equation}
 P=-2\textrm{Im}[i\partial_tc_{g}(t)c_{g}^{*}(t)]=\gamma _{g}(\varphi ). \label{rate}
\end{equation}
\noindent
It is possible to control different ionization products through the relative phase of the laser components.
The branching ratio into the $j$ channel is defined as
\(B_j= P_j/P\), where the partial ionization rates $ P_j{\rm }$ are:
\begin{eqnarray}
P_j&=&2\pi\left| \bar{D}_{gc_{j}}^{(4)}(t)+\bar{D}_{gc_{j}}^{(2)}(t)e^{2i\varphi }\right| ^{2},\quad j=1,2\\
P_3&=&2\pi \left|\bar{D}_{gc_{3}}^{(4)}(t)\right|^{2}.
\end{eqnarray}

Our calculation takes into account all the important resonances
with intermediate
atomic states for the process studied. For the range of intensities
where the two- and four-photon transition amplitudes are of comparable
values, quantum interference effects of the two ionization channels
would determine the modulation of the ionization signal. The photoionization
signal and the AI \( 3p^{2} \) line profile can be controlled by
varying the relative phase \( \varphi  \) between   the field components
of the laser field \( {\bf E}(t) \).

\noindent 
If we need to explicitly take into account the temporal
evolution of the laser pulse, we have to integrate the time-dependent
system of differential equations Eqs. (\ref{sys1})-(\ref{sys3}), with the initial
conditions \( c_{g}(t=0)=1 \) and \( c_{j}(t=0)=0,\, \, j=1,2,3 \).
The temporal pulse shape of the laser field is a sine-squared function, 
\( {\cal E}_{i}(t)={\cal E}_{i}\sin ^{2}(\pi t/T) \),
where \( i=f,h \). 
The integration time for this sine squared shape pulse is taken
from \( t=0 \) to 
 \( t=T=\frac{2\pi }{\omega _{f}}\textrm{n} \),
with \( \textrm{n} \) being the number of laser cycles. 

Additional parameters, such as the laser pulse duration and the temporal
overlap of the fundamental and harmonic pulse, may also control the
ionization signal.

The total ionization probability is calculated at the end of the laser
pulse: \( P_{ion}(T)=1-|c_{g}(T)|^{2} \). We have obtained the value
\( \Gamma _{3p^{2}}\simeq 1.76\times 10^{-3} \) a.u. for the autoionization
width of the \( 3p^{2} \) AIS, and the corresponding life-time of
this level is about \( \tau _{3p^{2}}=13.75 \) fs. The total laser
pulse duration used in our calculation is larger than the life-time
of the AIS.

\section{Atomic structure}

The Hamiltonian of the Magnesium atomic system \( H_{a} \) is given
by: \begin{equation}
\label{eq:hamiltonian}
H_{a}=\sum _{i=1}^{2}\left[ -\frac{1}{2}\nabla _{i}^{2}+V_{l}^{HF}(r_{i})\right] +\frac{1}{|{\mathbf{r}}_{1}-{\mathbf{r}}_{2}|}+V_{d}({\mathbf{r}}_{1},{\mathbf{r}}_{2}),
\end{equation}
 where \( V_{l}^{HF}(r_{i}) \) is the radial Hartree-Fock potential
for the closed-shell core of Magnesium (Mg\( ^{++} \)), and \( V_{d} \)
is a two-body interaction operator that includes a dielectronic effective
interaction \cite{chang:1993a,moccia:1996a}. The characteristic feature
of the Magnesium atom is the existence of a \( 3s^{2} \) valence
shell, outside a closed-shell core, the excitation of which requires
a much larger amount of energy compared with the first and second
ionization threshold. This allows us to explore excitation and/or
ionization processes of the valence electrons, for a certain range of photon energies, without considering the closed-shell core-excitation.

At a first stage, we perform a Hartree-Fock calculation for the closed-shell
core of Magnesium (Mg\( ^{++} \)), thus deriving  the effective Hartree-Fock
potentials 'seen' by the outer electrons. At a second stage, we solve
the Schr\"{o}dinger equation for the valence electrons through the
configuration interaction (CI) method using the Magnesium Hamiltonian
defined in Eq. (\ref{eq:hamiltonian}). In this case, we also add
a core polarization potential acting on the valence electrons, the
inclusion of which represents the influence of the core on the two
valence electrons in a way very similar to that described in \cite{chang:1993a,nikolopoulos:2003a}.
This core-polarization potential has the form 
\( \alpha _{s}\left\{ 1-\exp [-(r/r_{l})^{6}]\right\} /r^{4} \),
where \( \alpha _{s} \) is the static polarizability of the doubly-ionized
Mg, and  the cut-off radii, \( r_{l} \), for the various partial waves \( l=0,1,2,... \)
\cite{moccia:1996a}.

Having produced the Mg\( ^{+} \) one-electron radial eigenstates
\( P_{nl}(r) \) for each partial wave \( l=0,1,2,... \), we solve
the two-electron Schr\"{o}dinger equation: 
\begin{equation}
\label{se_{2}e}
H_{a}\Psi ^{\Lambda }({\textbf {r}}_{1},{\textbf {r}}_{2})=E\Psi ^{\Lambda }({\textbf {r}}_{1},{\textbf {r}}_{2}),
\end{equation}
 by expanding the two-electron eigenstates \( \Psi ^{\Lambda }({\textbf {r}}_{1},{\textbf {r}}_{2}) \)
on the basis of LS-uncoupled two-electron antisymmetrized orbitals
\( \Phi _{nln'l'}^{\Lambda }({\textbf {r}}_{1},{\textbf {r}}_{2}) \),
namely \cite{nikolopoulos:2003a}: 
\begin{equation}
\label{eq:wf_{2}e_{c}i}
\Psi ^{\Lambda }({\textbf {r}}_{1},{\textbf {r}}_{2};E_{i})=\sum _{nln'l'}C_{nln'l'}(E_{i})\Phi _{nln'l'}^{\Lambda }({\textbf {r}}_{1},{\textbf {r}}_{2}),
\end{equation}

\begin{equation}
\label{eq:wf_{2}e}
\Phi ^{\Lambda }_{nln'l'}({\textbf {r}}_{1,}{\textbf {r}}_{2})=A_{12}\frac{P_{nl}(r_{1})}{r_{1}}\frac{P_{n'l'}(r_{2})}{r_{2}}Y_{LM_{L}}(\hat{r}_{1},\hat{r}_{2};l,l'),
\end{equation}
 where \( \Lambda =(LM_{L}) \), and \( A_{12} \) is the antisymmetrization
operator which ensures that the total wave function is antisymmetric
with respect to interchange of the space coordinates of the two electrons.
Assuming  the Magnesium is  initially  in its ground state 
\( |3s^{2}\;^{1} S^e >\),
and that the transitions with a linearly polarized light are well described in the dipole approximation,
we only need to construct the singlet states with \( S=0 \)  and \( M_{L}=0 \).

We force the wavefunction to be zero at the boundaries by selecting
the basis functions \( P_{nl}(r) \), and \( P_{n'l'}(r) \) to be
the one-electron radial solutions of Mg\( ^{+} \) which, by construction
vanish at the boundaries. The radial functions \( P_{nl}(r) \) are
expanded in a B-spline basis of order \( n \), 
\( P_{nl}(r)=\sum _{i}b_{i}B_{i}(r),\, \, i=1,2,...,N \),
which transforms the one-electron Schr\"{o}dinger equation for the
Mg\( ^{+} \) orbitals into a system of matrix equations for the coefficients
\( b_{i} \) \cite{bachau:2001}.

Substitution of the two electron wavefunction Eq. (\ref{eq:wf_{2}e})
into the Schr\"{o}dinger equation Eq. (\ref{se_{2}e}), leads to a
generalized eigenvalue matrix equation, the diagonalization of
which gives the coefficients \( C_{nln'l'}(E_{i}) \) for each
discrete eigenvalue \( E_{i} \) \cite{nikolopoulos:2003a,lambropoulos:1998}.
This choice of the basis functions \( \Phi _{nln'l'}^{\Lambda }({\textbf {r}}_{1},{\textbf {r}}_{2}) \)
(constructed from the radial orbitals \( P_{nl}(r) \)),  thus leads
to a discretized continuum spectrum for the Magnesium atom with a density
of states basically determined  from the box radius \( R \).

Having obtained the two-electron wavefunctions 
\( \Phi _{nln'l'}^{\Lambda } \), 
 we calculate the effective two- and four-photon transition amplitudes
Eqs. (\ref{eq:cs2ph})-(\ref{eq:cs4ph}) within the lowest order-perturbation
theory (LOPT), as well as the corresponding  partial generalized \(N\)- photon cross
sections leading to the \(|c_j>\) continuum:
\begin{equation}
 \sigma ^{(N)}_{gc_{j}}(\omega )
=2\pi (2\pi \alpha )^{N}\omega ^{N}\left|{\cal M}_{gc_{j}}^{(N)}(\omega )\right|^{2}, \label{cs}
\end{equation} 
where \(\omega\) represents the laser frequency, \( \alpha  \) the fine structure constant, and
\(j=1,2\) for \( N=2\) (two-photon cross section) and \(j=1,2,3\) for \(N=4\) (four-photon cross section).
The effective two- and four-photons transition amplitudes are calculated, 
\(D_{gc_{j}}^{(N)}(\omega )=  I^{N/2}{\cal M}_{gc_{j}}^{(N)}(\omega )\), where $I $ represents the laser field intensity.
 We have employed a box with the radius close to 
\(R \approx 300\) a.u., 602 B-spline functions of order nine, and a total number of
 angular momenta up to \(L= 4\).

\section{Results and discussion}
\noindent
In the first part of this section we present a few results in the weak field limit 
for a time-independent laser pulse,
in which case the ionization rate approximation is valid, and the photoionization
signal is  well described by the ionization rate formula Eq. (\ref{rate}).

\begin{figure}
\centerline{\resizebox*{8cm}{!}{\includegraphics{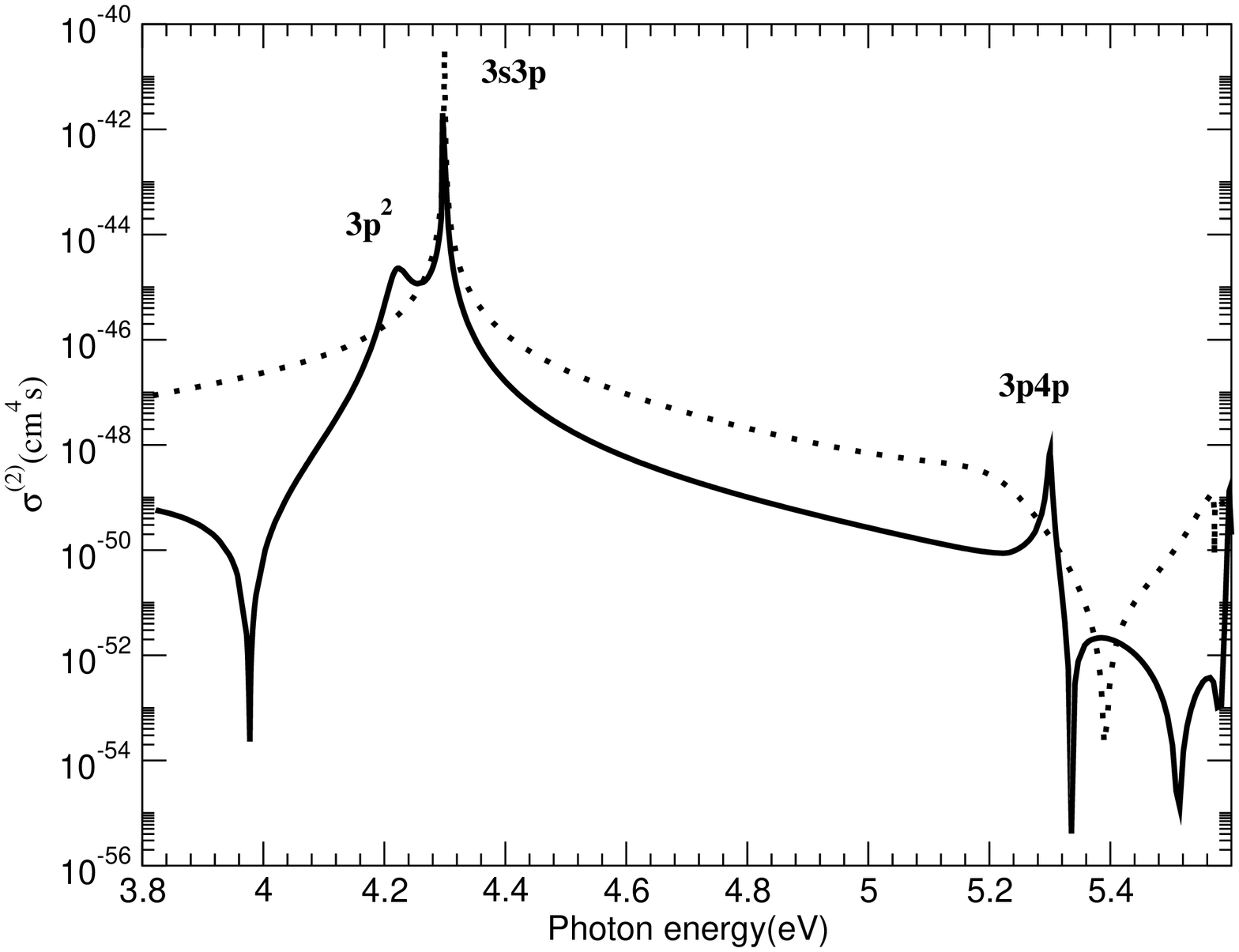}} }
\caption{\label{fig:fig2}}\vspace*{1cm}
\end{figure}

\begin{figure}
\centerline{\resizebox*{8cm}{!}{\includegraphics{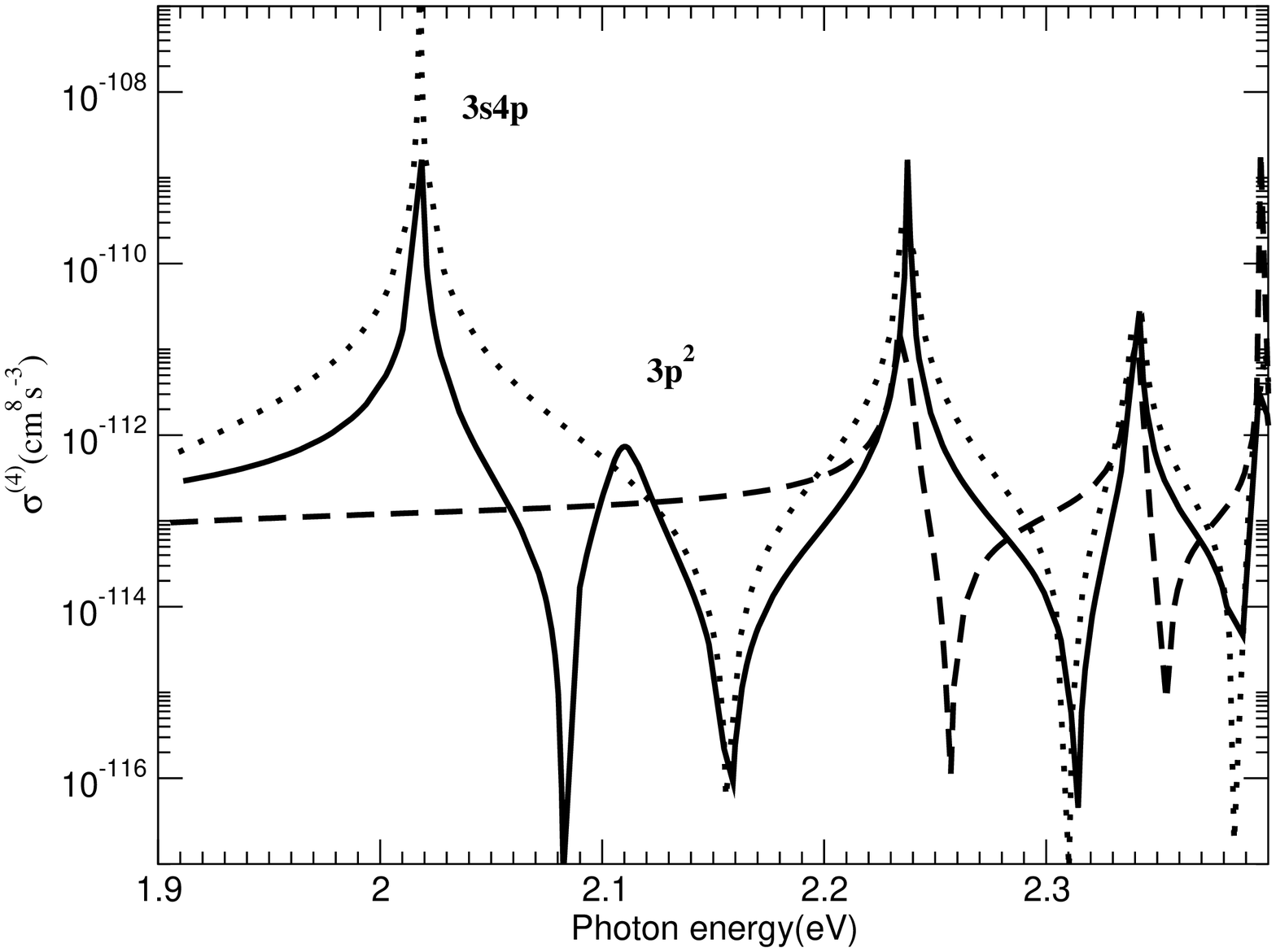}} } 
\caption{\label{fig:fig3}} \vspace*{1cm}
\end{figure}

Figure \ref{fig:fig2} shows the partial two-photon ionization cross sections
from the ground state of the Magnesium atom leading to the
\( ^{1}S^e \) (full curve)
 and \( ^{1}D^e \) continuum (dotted curve), as a function of the photon energy.
 Our results presented in Figure \ref{fig:fig2} are in agreement with the
results published in the literature by Chang and Tang \cite{chang:1992} and
Kylstra \textit{et al.} \cite{kylstra:1998}.
The four-photon partial cross sections from the ground state leading to the  \( ^{1}S^e \)
(full curve),  \( ^{1}D^e \) (dotted curve), and 
\( ^{1}G^e \) (dashed curve) continua are presented
 in Figure \ref{fig:fig3}. The two- and four-photon
partial cross sections were calculated using Eq. (\ref{cs}). 

The total ionization rate for the simultaneous two- and four-photon
ionization from the ground state of Mg is plotted as a function of
the fundamental laser frequency in Figure \ref{fig:fig4}, for a relative
phase \( \varphi =0^{0} \) (dashed line),  \( 30^{0} \) (dotted line), 
and \( 90^{0} \) (full line). The laser field intensities
are \( I_{f}=2\times 10^{11} \) W/cm\( ^{2} \) and \( I_{h}=4.13\times 10^{7} \)
W/cm\( ^{2} \). If the two fields are in phase, there is destructive
interference between the two different ionization channels illustrated
in Figure \ref{fig:fig1}, while there is constructive interference at \( \varphi =90^{0} \).
This behavior of the ionization rate is due to the relative sign of the  
effective two- and four-photon transition amplitudes;
specifically, for a fundamental laser frequency at
 2.08 eV$< \omega_f <$ 2.15 eV, the effective two- and four-photon transition amplitudes
from the ground state into 
the \( ^{1}S^e \) and  \( ^{1}D^e \) continua, respectively, have opposite signs.
 When the laser
frequency is tuned near the resonant state \( 3s3p \) (\( \omega _{h} \approx 4.3 \),
eV) no phase effects are observed, since the ionization through two harmonic
photon absorption becomes dominant over the four photon absorption,
Figure \ref{fig:fig1}(ii), and no interference process exists. Where
the laser is tuned around the \( 3s4p \) state (\( \omega _{f}\approx 2.02 \)
eV), channel (i) of Figure \ref{fig:fig1} is dominant.

In order to have a better view of the interference process between the two
ionization channels illustrated in Figure \ref{fig:fig1}, in Figure \ref{fig:fig5} we present 
the partial ionization rate into the \( ^{1}S^e \) (dot-dashed curve),
\( ^{1}D^e \)(dotted curve) and   \( ^{1}G^e \) continua (dashed curve).
The full curve represents the total ionization rate when the fields are in phase, 
 panel (a), and when the relative phase is  \( \varphi =90^{0} \),  panel (b).
We observe that the AI \( 3p^{2} \) line profile for the two- and four- photon
process is not symmetric due to the one- and three-photon transitions to
the \( 3s3p \) and \( 3s4p \) near resonance states.

Figure \ref{fig:fig10}(a) shows the branching ratios function of the fundamental laser
frequency at \( \varphi =0^{o} \), and 
in panel b) for \( \varphi =90^{o} \) with the same conditions as in Figure \ref{fig:fig4}.
 The branching ratio into the  \( ^{1}S^e \) continuum is
 represented  by the dot-dashed curve, into the  \( ^{1}D^e \)
 continuum by  the dotted curve and into the   \( ^{1}G^e \) continuum
 by the dashed curve. 
As shown in panel (a),  at \( \varphi =0^{0} \)
 for laser frequencies tuned around  $3p^2$ AIS, the ionization rate
 into the \( ^{1}G^e \) continuum is enhanced since the laser intensities
 where chosen such that the two- and four-photon transition amplitudes 
from the ground state into  the \( ^{1}S^e \) continuum  cancel each
 other.
On the other hand, at  \( \varphi =90^{0} \) in panel (b) there is an enhancement of the  ionization signal into the \( ^{1}S^e \) and \( ^{1}D^e \)  continua.
This suggests  that by a judicious choice of the laser
 intensities,  it might be possible to arrange the maximum of one ionization product
to coincide with the minimum of the others, and thus to control different ionization products.

The modulation of the ionization rate as a function of the relative phase for a
simultaneous two- and four-photon ionization from the ground state
of Mg to the $3p^2 (^{1}S^e)$ AIS  is depicted in Figure \ref{fig:fig6}, for \( \omega _{f}=2.11 \)
eV with the same laser intensities as in Figure \ref{fig:fig4}. The full curve
represents the total ionization rate, the dot-dashed curve is the contribution
to ionization into the \( ^{1}S^e \) continuum, and the dotted curve
the contribution coming from the \( ^{1}D^e \) continuum. 
 The ionization rate into the \( ^{1}G^e \) continuum (dashed curve) is a flat
line and contributes  to the background of the ionization rate, which it cannot
 be neglected when the relative phase between the laser field components is close to \( k\pi \).
The ionization rate  into the  \( ^{1}S^e \) at \( \varphi =k\pi \) is zero since the laser intensities
were chosen such that the effective transition amplitudes for two-
and four-photon absorption into the \( ^{1}S^e \) continuum  at  \( \omega _{f}=2.11 \) eV are almost
equal, and cancel each other.

 We can analyze the ionization rate as a function of the harmonic
  intensity for fixed frequency and intensity of the fundamental.
In Figure \ref{fig:fig7} we show  the depth of ionization \(
(P_{max}-P_{min})/\frac{1}{2}(P_{max}+P_{min}) \) as a function of the harmonic
intensity. \( P_{max} \) represents the maximal value of the ionization rate when
\(\varphi=90^{0}\), and \( P_{min} \) is the minimal value of the ionization rate
when the two laser components are in phase \(\varphi=0^{0}\); the fundamental
photon energy is $\omega_f=2.11$ eV.
As can be seen, an efficient coherent control in the weak field limit is obtained
at \( I_{f}=2\times 10^{11} \) W/cm\( ^{2} \) for the harmonic intensity
in the interval \( I_{h}\in (2\times10 ^{7} \)W/cm\( ^{2} \), \( 2\times 10^{8} \)
W/cm\( ^{2} \)).
 Of course, good coherent control can be obtained as well for other pairs of laser  intensities.

When  considering a temporal dependence of the laser pulse shape, the total
ionization probability is obtained by numerically integrating  the system of 
differential equations for the amplitude coefficients in
Eqs. (\ref{sys1})-(\ref{sys3}).
In Figure \ref{fig:fig8} we  plot the ionization yield
for  field intensities where the depth of modulation has a maximal value:
 \( I_{f}=2\times 10^{11} \) W/cm\( ^{2} \)
and \( I_{h}=5\times 10^{7} \) W/cm\( ^{2} \). The ionization
yield for the relative phase \( \varphi =0^{0} \) is described by
the dashed  curve, for \( \varphi =30^{0} \) by the dotted curve, and
for \( \varphi =90^{0} \) by the full curve. The laser pulse shape
is sinusoidal with a total duration of 1 ps (\( {\textrm{n}} \)=500).

Figures \ref{fig:fig4}-\ref{fig:fig6} and \ref{fig:fig8}
 clearly show that the AI
line shape of \( 3p^{2} \) is significantly changed by the relative
phase \( \varphi  \) of the laser components. The corresponding
\( 3p^{2} \) peak at \( \omega _{f}=2.11 \) eV is  diminished the most
whenever the two fields are in or out of phase due to the destructive
quantum interference between the two different channels, and enhanced the most
whenever \( \varphi =(2k+1)\pi /2 \) due to the constructive  interference
of the two ionization pathways. The maximum and the minimum of the
ionization signal differ by more than one order of magnitude.

In order to better illustrate  the destructive interference between
the two- and four-photon ionization processes,  in Figure \ref{fig:fig9} we plot
the ionization yield as a function of the fundamental laser frequency
at \( \varphi =0^{0} \), and the same laser intensities as in Figure \ref{fig:fig8}.
The laser pulse shape is sinusoidal with a total duration of about
\( 1 \) ps.
The dashed line represents the ionization rate corresponding to
the four-photon absorption from the fundamental, and the dotted line
describes the ionization yield due to the two-photon absorption of
its second harmonic. The full line is the total ionization yield resulting
from the destructive interference of these two processes.

\section{Conclusion}

In this paper, we have investigated the phase  coherent effect of the $3p^2$ ($^1S^e$)
autoionizing state  resonantly coupled to the  ground state of a 
  Mg atom  in the presence  of a bichromatic laser field of frequencies $\omega_f$
 and $2\omega_f$. The transition amplitudes have been
evaluated using  a realistic atomic structure calculation.
The motivation for this study was the investigation of the possibility of 
achieving  coherent control of the photoelectron current and   
changes in the autoionization line shape. 
We have calculated and presented, for the first time (to the best of our knowledge),
  four-photon partial transition amplitudes for the Mg atomic system, and 
the corresponding partial  ionization cross sections. 
 We have found out that the near resonance, double excited states $ 3s3p\; (^1P^{o}$) and
 $3s4p\;(^1P^{o}$) of Magnesium introduce distortions in the left and
  in the right side  of the autoionizing resonance,
 and enhance the overall ionization signal.
A reliable study of the \( 3p^{2} \) line shape through the density
 matrix or resolvent operator method
should take into consideration these two intermediate resonances, and the energy
 dependence of the respective transition amplitudes and Rabi frequencies.
 The relative phase $\varphi$ between the two components of the 
bichromatic laser field  modulates the quantum interference between the 
two ionization channels: four-photon absorption  from the fundamental
laser field and two-photon absorption of its second harmonic.

\begin{acknowledgement} The work by G.B-Z. was supported by the European
Research Network Program Contract No HPRN-CT-1999-00129. One of the
authors (G.B-Z.) is indebted to Professor P. Lambropoulos for useful discussions. 
\end{acknowledgement}

\newpage
{\centering TABLE - FIGURE CAPTIONS }

Fig.  1. Energy diagram of the Mg atom interacting with a bichromatic
electromagnetic field, showing the relevant levels for the studied process.

Fig.  2. Two-photon ionization cross section \( \sigma ^{(2)} \), leading to 
\(^{1}S^e\) (full line)  and \(^{1}D^e\) (dotted line) continua  from
the ground state of Mg,   as a function of
the photoelectron energy.

Fig.  3. Four-photon generalized ionization cross section \( \sigma ^{(4)} \),
leading to \(^{1}S^e\) (full line), \(^{1}D^e\) (dotted line)  and
\(^{1}G^e\) (dashed line) continua from the ground 
state of Mg, as a function of the photoelectron energy.

Fig.  4. The ionization rate of the ground state of Mg as a function
of the fundamental laser frequency for  \( \varphi =0^{0} \) (dashed
line),  \( 30^{0} \) (dotted line), and \( 90^{0} \) (full line).
 The peak occurring at 2.02 eV
corresponds  to the \( 3s4p \) level, and the peak at
2.15 eV corresponds to the \( 3s3p \) level. The fundamental and
harmonic laser intensity are \( I_{f}=2\times10 ^{11} \) W/cm\( ^{2} \)
and \( I_{h}=4.13\times10 ^{7} \) W/cm\( ^{2} \), respectively.

\begin{figure}
\centerline{\resizebox*{8cm}{!}{\includegraphics{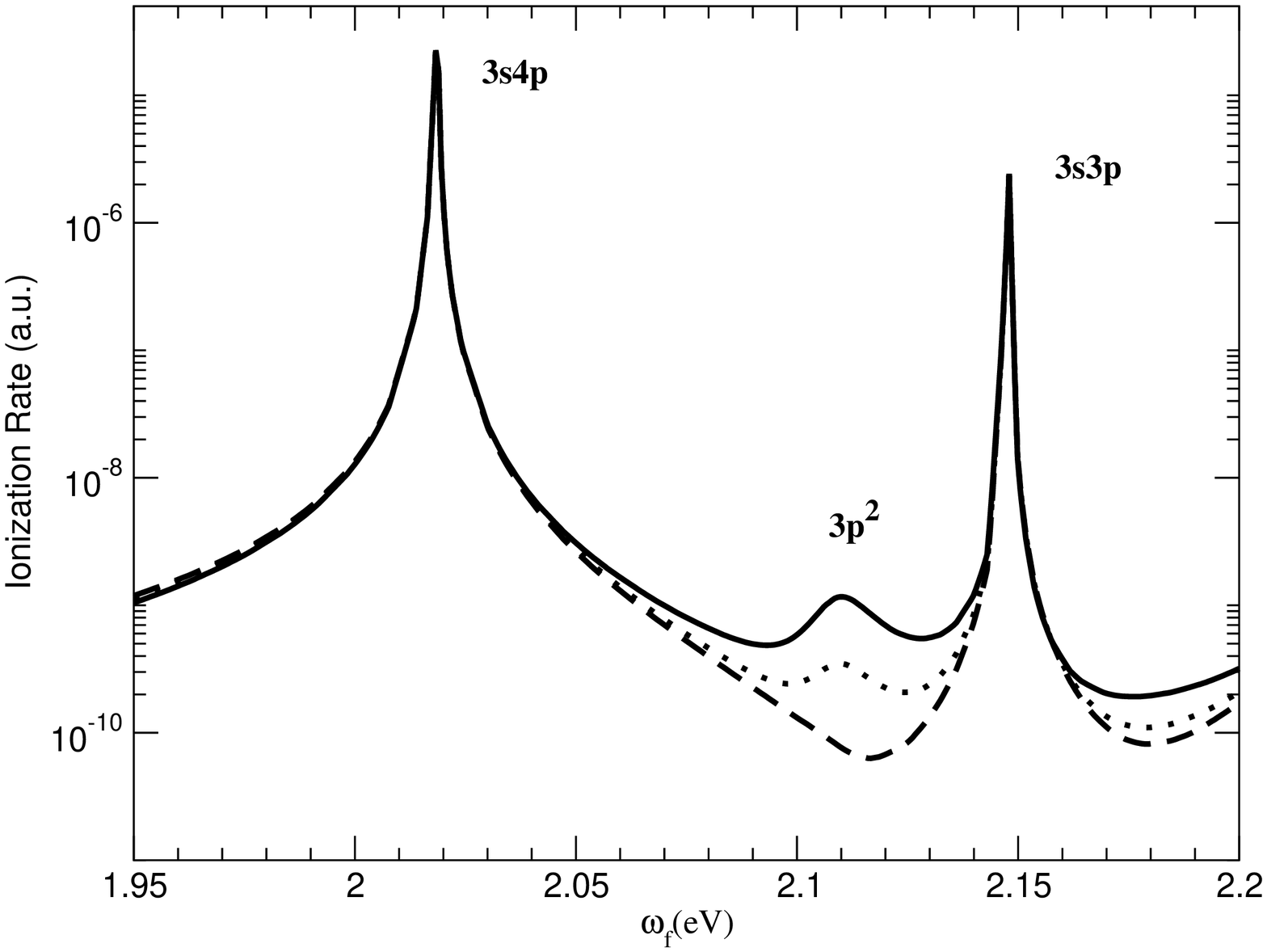}} }
\caption{\label{fig:fig4}} \vspace*{1cm}
\end{figure}

\begin{figure}
\centerline{\resizebox*{8cm}{!}{\includegraphics{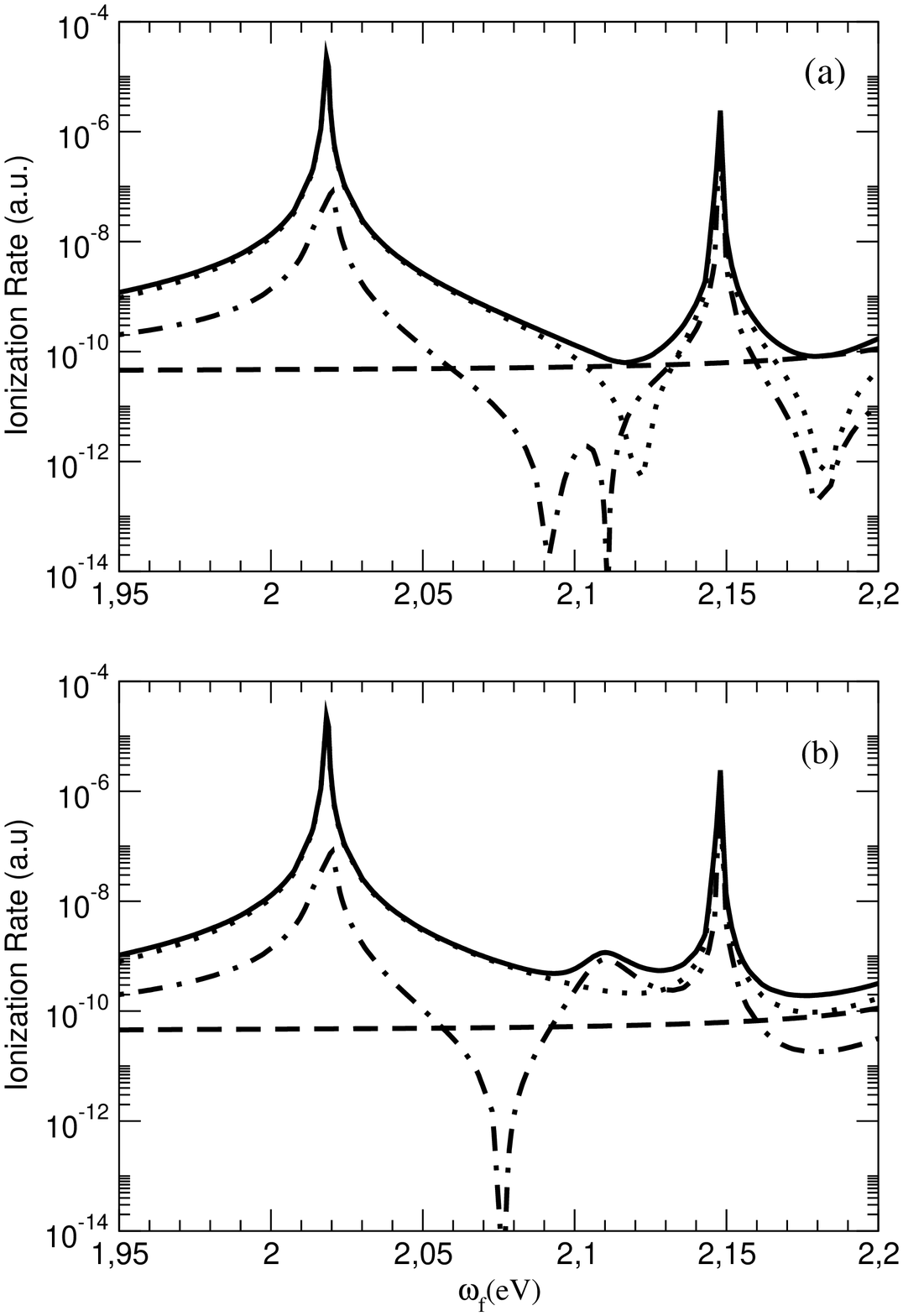}} } 
\caption{\label{fig:fig5}}\vspace*{1cm}
\end{figure}

\begin{figure}
\centerline{\resizebox*{8cm}{!}{\includegraphics{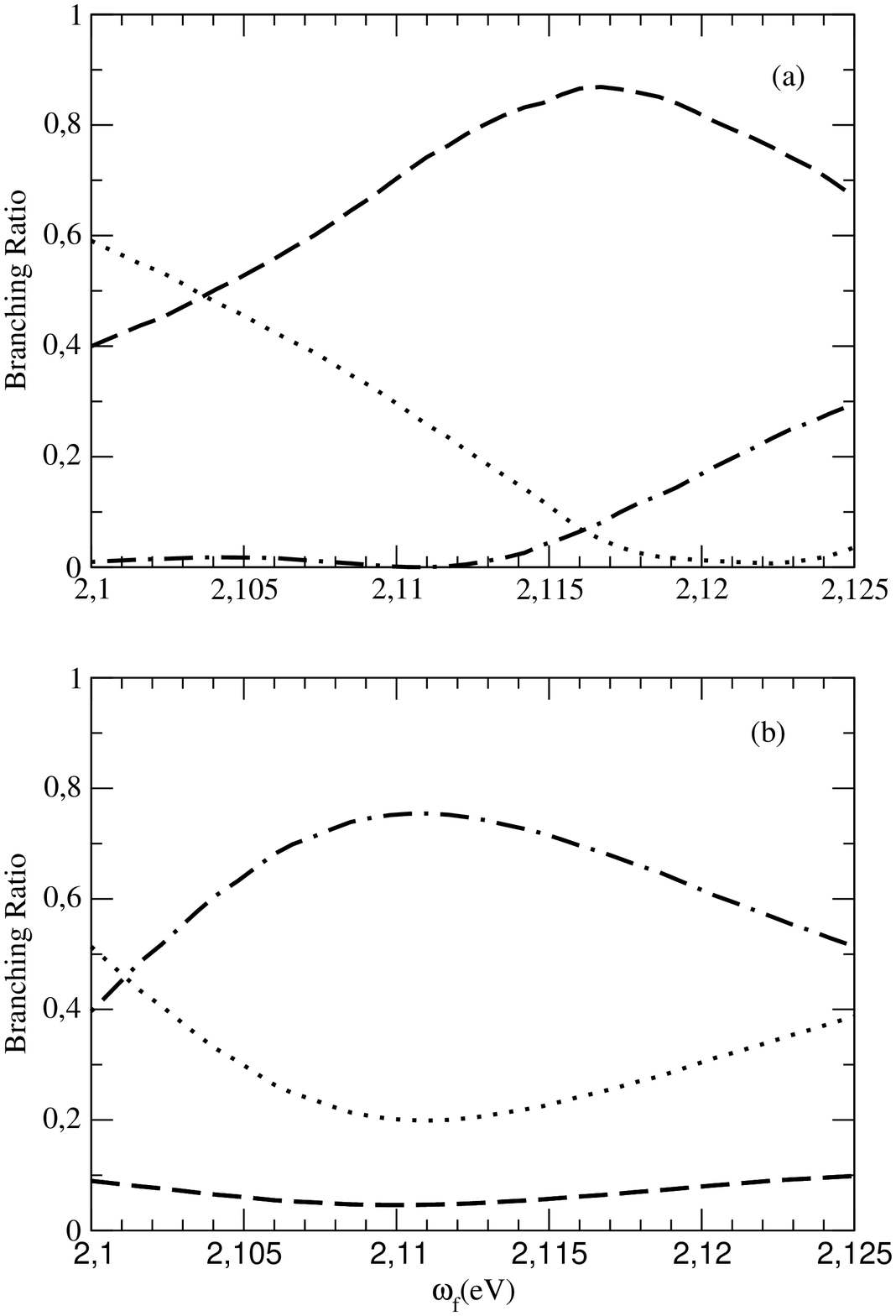}} } 
\caption{\label{fig:fig10}}\vspace*{1cm}
\end{figure}

\begin{figure}
\centerline{\resizebox*{8cm}{!}{\includegraphics{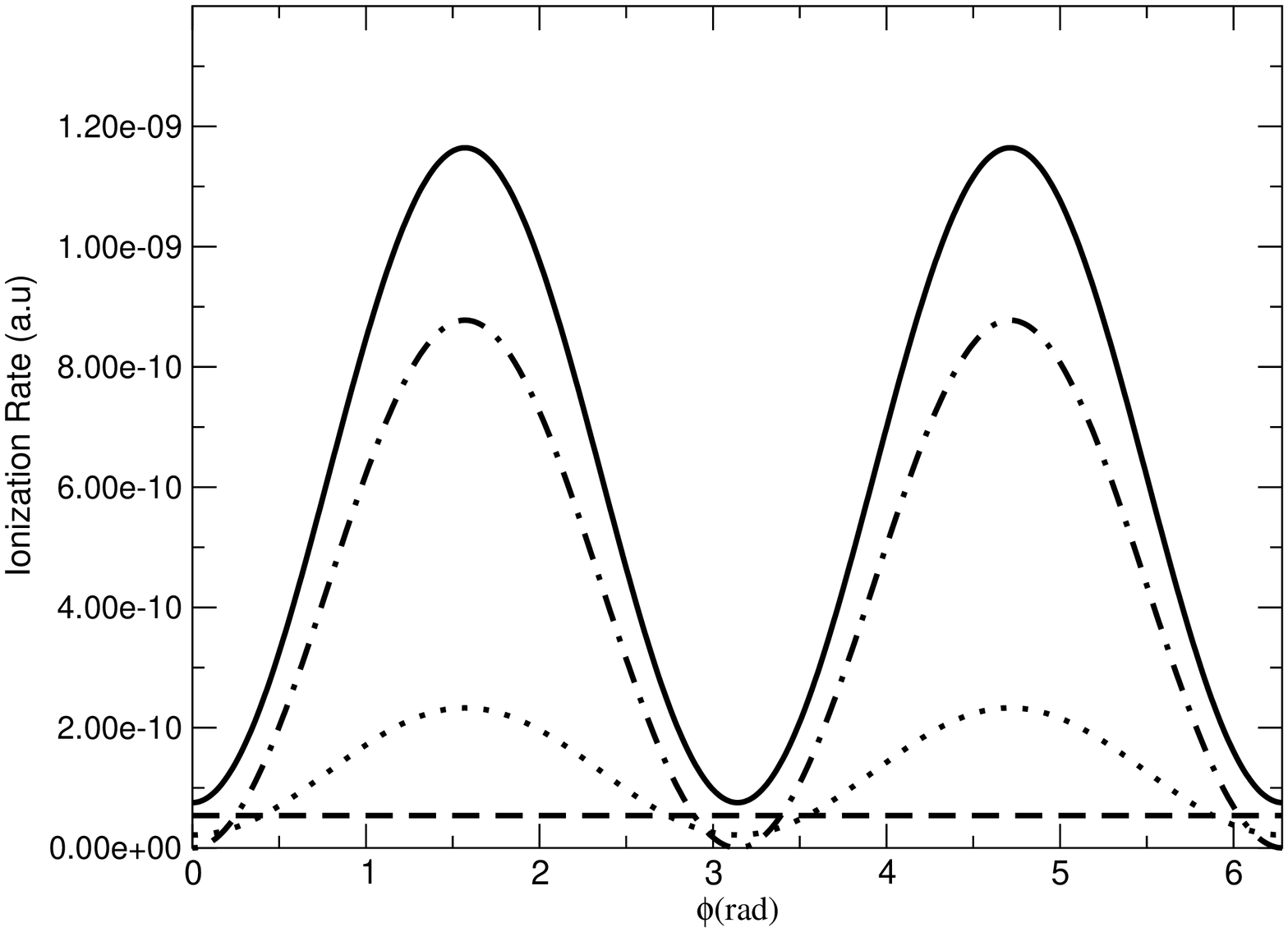}} } 
\caption{\label{fig:fig6}}\vspace*{1cm}
\end{figure}

\begin{figure}
\centerline{\resizebox*{8cm}{!}{\includegraphics{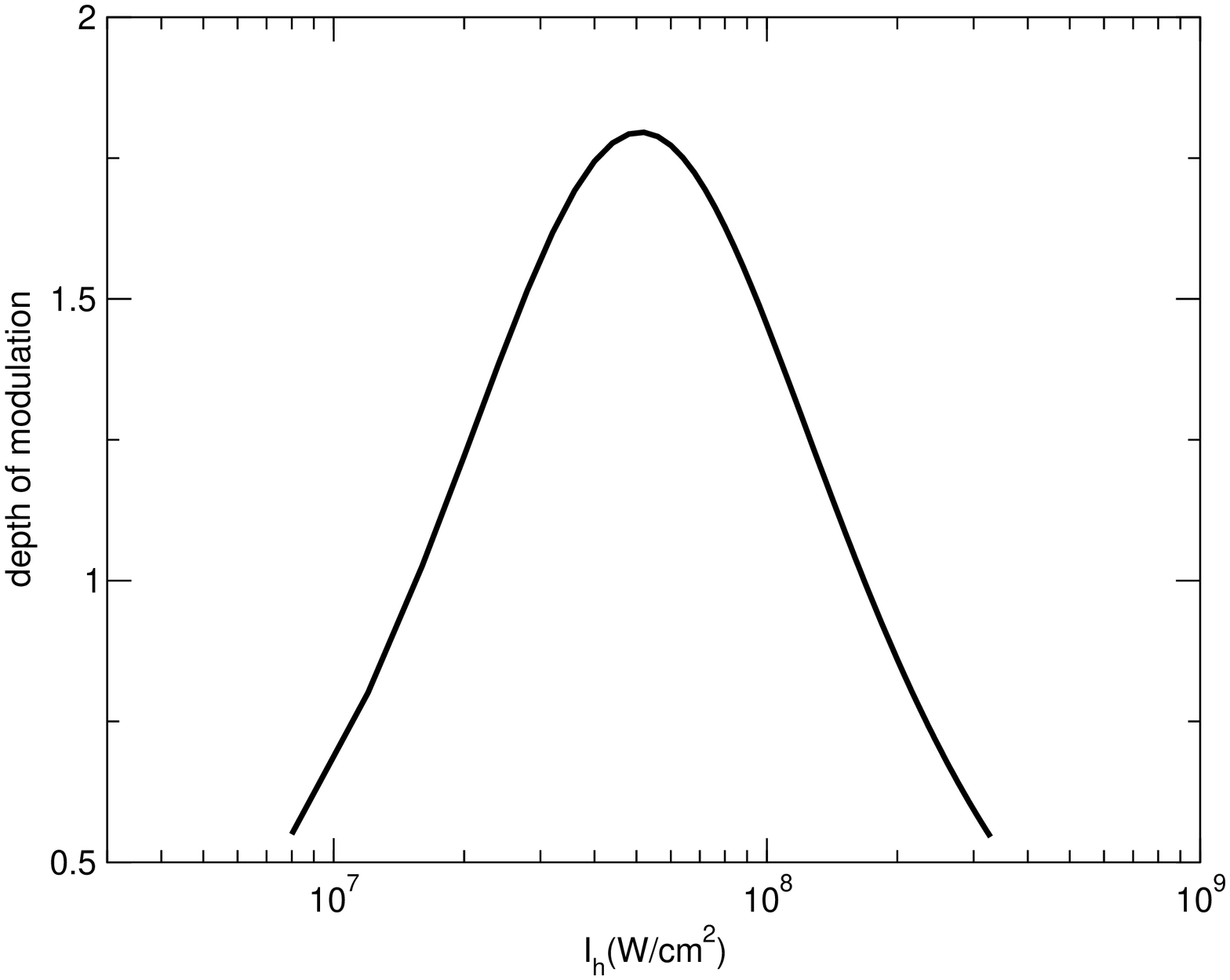}} }
\caption{\label{fig:fig7}}\vspace*{1cm}
\end{figure}

Fig.  5. The total and the partial ionization rates  of the ground state
 of Mg  as a function of the fundamental laser frequency.
 In  panel (a) the relative phase between field components
is $0^0$, and is  $90^0$ in  panel (b).
The partial ionization rate into the  \( ^{1}S^e \) continuum is
 represented  by the dot-dashed curve, into the  \( ^{1}D^e \)
 continuum by  the dotted curve and into the   \( ^{1}G^e \) continuum
 by the dashed curve. The full line represents the total ionization
 rate.
The laser parameters are the same as in Figure  \ref{fig:fig4}.

Fig. 6. Branching ratios as a  function of the fundamental laser
frequency at \( \varphi =0^{o} \) in panel (a), and at \( \varphi =90^{o} \)
in panel (b).
 The branching ratio into the  \( ^{1}S^e \) continuum is
 represented  by the dot-dashed curve, into the  \( ^{1}D^e \)
 continuum by  the dotted curve, and into the   \( ^{1}G^e \) continuum
 by the dashed curve. 
The laser parameters are the same as in Figure  \ref{fig:fig4}.

Fig. 7. Ionization rate vs the relative phase between laser components.
The laser intensities are the same as in Figure \ref{fig:fig4}.
The partial ionization rate into the  \( ^{1}S^e \) continuum is
 represented  by the dot-dashed curve, into the  \( ^{1}D^e \)
 continuum by  the dotted curve and into the   \( ^{1}G^e \) continuum
 by the dashed curve. The full line represents the total ionization rate.

Fig. 8. The depth of modulation as a function of the harmonic intensity.
The fundamental laser intensity is \( I_{f}=2\times10 ^{11} \) W/cm\( ^{2} \).

Fig. 9. The ionization yield of the ground state of Mg around the \( 3p^{2} \)
AIS  function of the fundamental laser frequency for  \( \varphi =0^{0} \) (dashed
line),  \( 30^{0} \) (dotted line), and \( 90^{0} \) (full line).
 The fundamental and harmonic peak laser intensity are \( I_{f}=2\times10 ^{11} \)
W/cm\( ^{2} \) and \( I_{h}=5\times10 ^{7} \) W/cm\( ^{2} \),
respectively. The laser pulse shape is sinusoidal with a total duration
of \( 1 \) ps.

\begin{figure}
\centerline{\resizebox*{8cm}{!}{\includegraphics{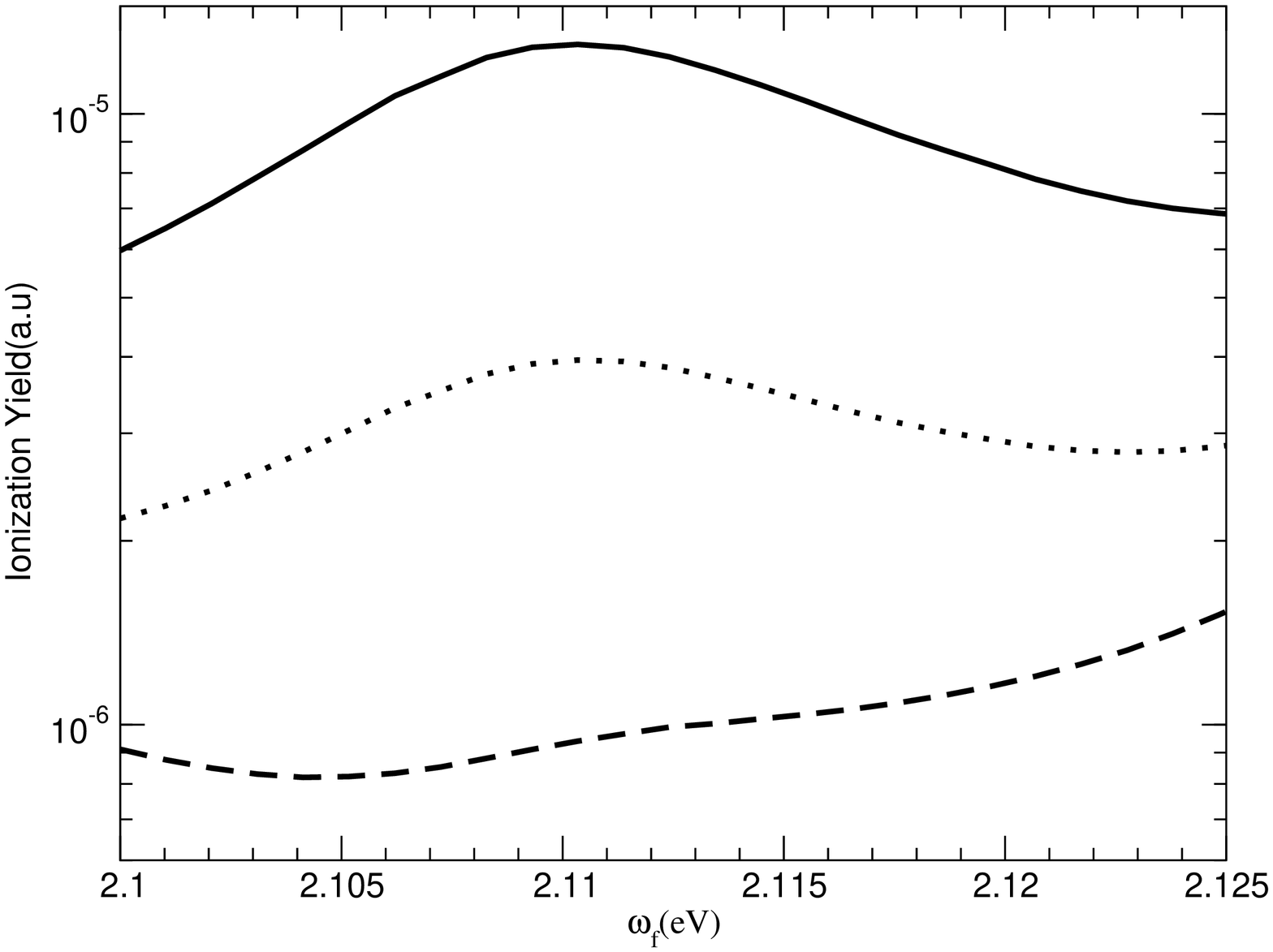}} } 
\caption{\label{fig:fig8}}\vspace*{1cm}
\end{figure}

\begin{figure}
\centerline{\resizebox*{8cm}{!}{\includegraphics{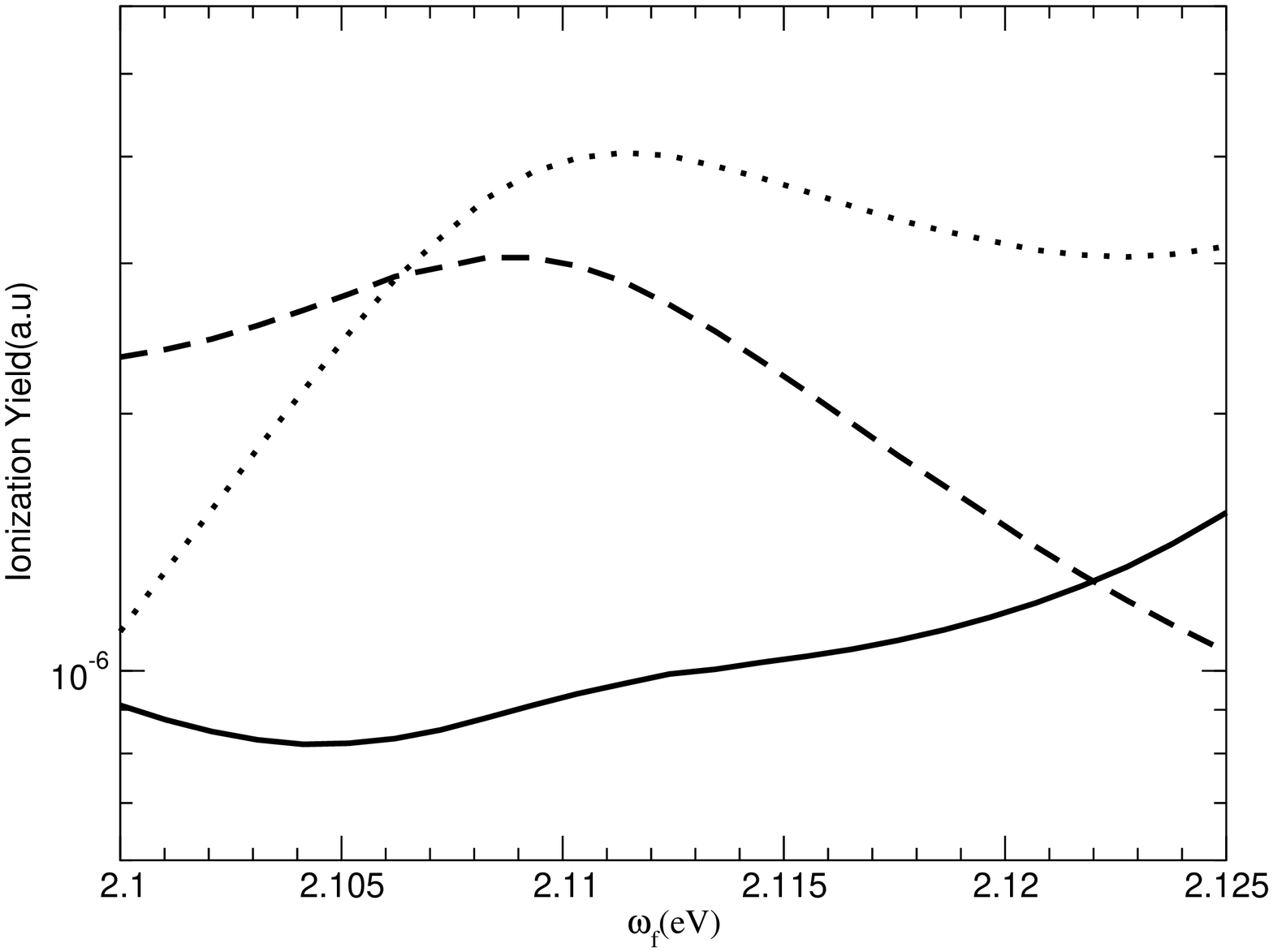}} } 
\caption{\label{fig:fig9}}\vspace*{1cm}
\end{figure}

Fig. 10. The ionization yield at \( \varphi =0^{o} \) as a function of the
fundamental laser frequency (full curve).
 The dotted curve represents the ionization yield for the two-photon absorption,
 and the dashed one corresponds to the four-photon absorption. The peak
laser intensities are the same as in Figure \ref{fig:fig8}.

\newpage

\end{document}